\documentclass{moriond}

\def\Journal#1#2#3#4{{#1} {\bf #2}, #3 (#4)}

\def\PLB{{\em Phys. Lett.}  B}

\def\PRD{{\em Phys. Rev.} D}
\def\JINST{\em Journal of Instrumentation}

\def\JHEP{{\em JHEP.}}
\def\EPJC{{\em EPJC.}}

\def\be{\begin{equation}}
\def\ee{\end{equation}}
\def\bea{\begin{eqnarray}}
\def\eea{\end{eqnarray}}

\def\hgg{\ensuremath{H \rightarrow \gamma\gamma}~}
\def\hzz{\ensuremath{H \rightarrow Z \rightarrow 4l}~}
\def\hww{\ensuremath{H \rightarrow WW}~}
\def\htt{\ensuremath{H \rightarrow \tau\tau}~}
\def\hbb{\ensuremath{H \rightarrow bb}~}
\def\pt{\ensuremath{p_{\mathrm{T}}}~}
\def\ptn{\ensuremath{p_{\mathrm{T}}}}

\newcommand{\Photo}{}

\begin{document}
\vspace*{4cm}
\title{Measurement of the differential Higgs boson production cross sections in ATLAS and CMS experiments}

\author{ Abdollah Mohammadi on behalf of ATLAS and CMS collaborations }

\address{Department of Physics, University of Wisconsin-Madison, \\
Madison, Wisconsin 53706, USA}

\maketitle\abstracts{
A comprehensive set of studies has been conducted to measure the fiducial differential cross sections of Higgs boson production by the ATLAS and CMS experiments. These analyses are based on proton-proton collision data collected by both experiments at the CERN LHC, with a center-of-mass energy of 13 TeV. The data sample corresponds to an integrated luminosity of up to 140 fb$^{-1}$. Fiducial Differential cross sections are measured for various observables sensitive to the production and decay of the Higgs boson. All measurements align with Standard Model predictions. The results are used to constrain potential anomalous interactions between the Higgs boson and other Standard Model particles.
}

\section{Introduction}

The discovery of the Higgs boson in 2012 by the ATLAS and CMS experiments at the Large Hadron Collider (LHC) \footnote{Copyright 2022 CERN for the benefit of the ATLAS and CMS Collaborations. Reproduction of this article or parts of it is allowed as specified in the CC-BY-4.0 license} marked a monumental milestone in the field of particle physics~\cite{ATLAS:dis,CMS:dis,CMS:dis2,ATLAS:det,CMS:det}. This discovery not only confirmed the existence of the last missing elementary particle predicted by the Standard Model (SM) but also provided crucial insights into the mechanism of electroweak symmetry breaking and mass generation for elementary particles.
Since its discovery, significant effort has been devoted to studying the properties of the Higgs boson in detail. One of the key aspects of these studies is the measurement of the Higgs boson production cross sections. These measurements are critical for testing the predictions of the SM and for searching for possible signs of new physics beyond the SM.
The differential cross section of the Higgs boson provides a more granular view of its production mechanisms and decay channels. Unlike the total cross section, which gives an overall production rate, the differential cross section describes how the production rate varies with respect to kinematic variables such as the Higgs boson transverse momentum and pseudorapidity, jet multiplicity, and other variables.
Analyzing these distributions allows us to probe the underlying dynamics of Higgs production and to identify deviations from the SM predictions. Such deviations could indicate the presence of new particles or interactions that modify the Higgs boson's properties.

\section{Analysis strategy}

Fiducial differential measurements represent the most model-independent way to measure Higgs boson production cross section. 
The differential cross sections, such as measured as a function of the Higgs boson \ptn, can be distorted by variations in the Higgs boson couplings
Therefore, differential cross section distributions can be sensitive to the existence of beyond-the-SM effects. In s fiducial measurement the region of interest is aligned with detector configuration; this minimizes the assumptions for extrapolation to full phase-space.


A similar strategy has been adopted to measure the differential production cross section in all Higgs boson decay modes.
After event selection and background estimation are completed, the analysis flow for conducting the differential cross section measurement proceeds as follows:
\begin{enumerate}
\item {\it Reconstructing an observable:} This step is purely analysis-dependent, with the goal of choosing the observable that separate the Higgs boson signal from SM backgrounds. The Higgs boson signal is extracted using a maximum likelihood fit to the
chosen observable distribution. The choice of the observable is generally consistent across both experiments, and tabulated in Table~\ref{tab:observables}.
\item {\it Variable definition:} 
In this step, the variables used to parameterize the Higgs boson cross section are defined. The most common variable sensitive to deviations from the Standard Model is the Higgs boson \ptn. Other variables, such as the \pt of the leading jet, the $\eta$ of the Higgs boson, and jet multiplicity, have also been considered. Given a sufficiently large data sample, 2-dimensional differential cross sections have been measured as well.
\item {\it Unfolding method:}  
To derive the differential production cross section, the Higgs boson signals are split into several bins based on the value of the generated- (gen-)level variable.
The same binning is used at reconstructed- (reco-) level  to categorize events.
The gen- and reco-level observable values are not perfectly aligned 
because of the limited resolution, causing  some events from one gen-level bin to migrate to  
a different reco-level bin. The unfolding method is used to correct for these effects. This is done by help of a response matrix that describes the probability of measuring the cross section in a reco bin given the gen value. The detector response matrix accounts for bin-to-bin migrations in the unfolding of the signal.
\item {\it Fitting procedure:} By performing one simultaneous fit over all reco-level bins of the chosen variables, the signal strength modifiers, defined as the ratio of the observed signal yield to the expected yield predicted by the SM, of the gen-level
observable bins are determined utilizing the full statistical power of the data set. The systematic uncertainties are treated as nuisance parameters. The measured signal strength modifier for each bin of variable is used to plot differential cross section production distribution. 
\item {\it Interpretation:} The last step is the interpretation of the results in Standard Model Effective Field Theory (SMEFT), $\kappa-$framework, or other frameworks and will be discussed in more detail in section~\ref{sec:SMEFT}.
\end{enumerate}



\begin{table}[http]
\caption{List of Higgs boson decay modes and the relevant observables used in the maximum likelihood fit to measure the differential cross section for each channel and experiment}
\begin{center}
\begin{tabular}{|c|c|c|}
\hline
Channel/Experiment & ATLAS  &CMS \\
\hline
\hgg & $m_{\gamma\gamma}$[~\cite{ATLAS:hgg}] & $m_{\gamma\gamma}$[~\cite{CMS:hgg}] \\
\hline
\hzz & $m_{4l}$ [~\cite{ATLAS:hzz}] &  $m_{4l}$ [~\cite{CMS:hzz}]\\
\hline
\hww (inclusive \& VBF) & $m_{ll}$[~\cite{ATLAS:hww,ATLAS:hwwVBF}] &  $m_{ll}$[~\cite{CMS:hww}] \\
\hline
\hbb & $m_{bb}$[~\cite{ATLAS:hbb,ATLAS:hbbWZ}]  &  $m_{bb}$[~\cite{CMS:hbb}]\\
\hline
\htt (non-boosted \& boosted) & - &  $m_{\tau\tau}$[~\cite{CMS:htt}]   \& NN[~\cite{CMS:httboost}]   \\
\hline
Combined &  [~\cite{ATLAS:cmb}]  & [~\cite{CMS:cmb}]   \\
\hline
\end{tabular}
\end{center}
\label{tab:observables}
\end{table}%



\section{Fiducial differential cross section results}

The fiducial differential cross section have been measured in all five main Higgs boson decay channels. The \hgg channel represents a channel with excellent diphoton mass resolution. 
Signal peaks at a mass of around 125 GeV over a smoothly falling background, mainly composed of diphoton continuum and photon+jet backgrounds. The total background is estimated from data using diphoton mass sideband regions.  

\begin{figure}[http]
\centering
\includegraphics[width=0.35\linewidth]{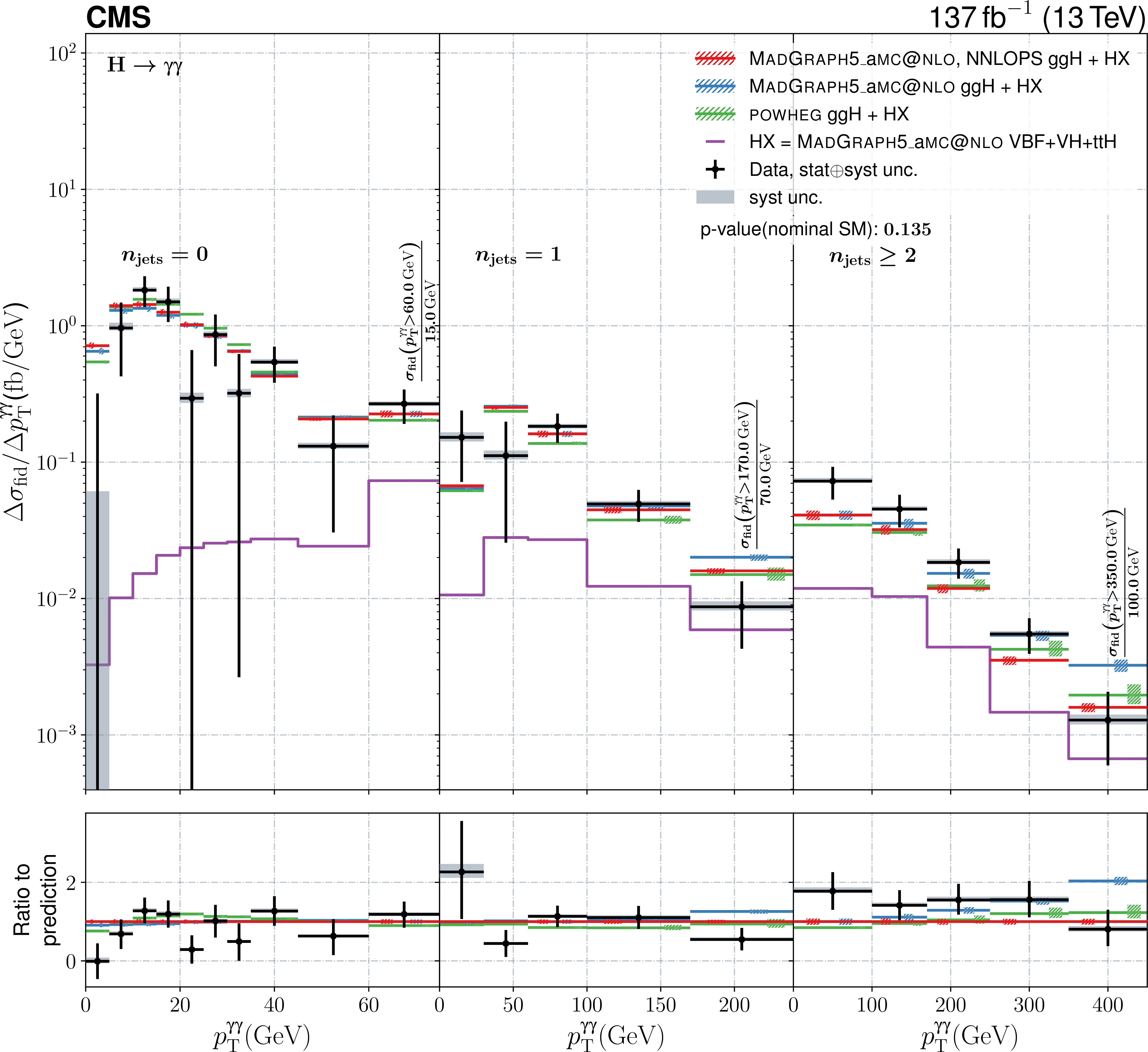}
\includegraphics[width=0.34\linewidth]{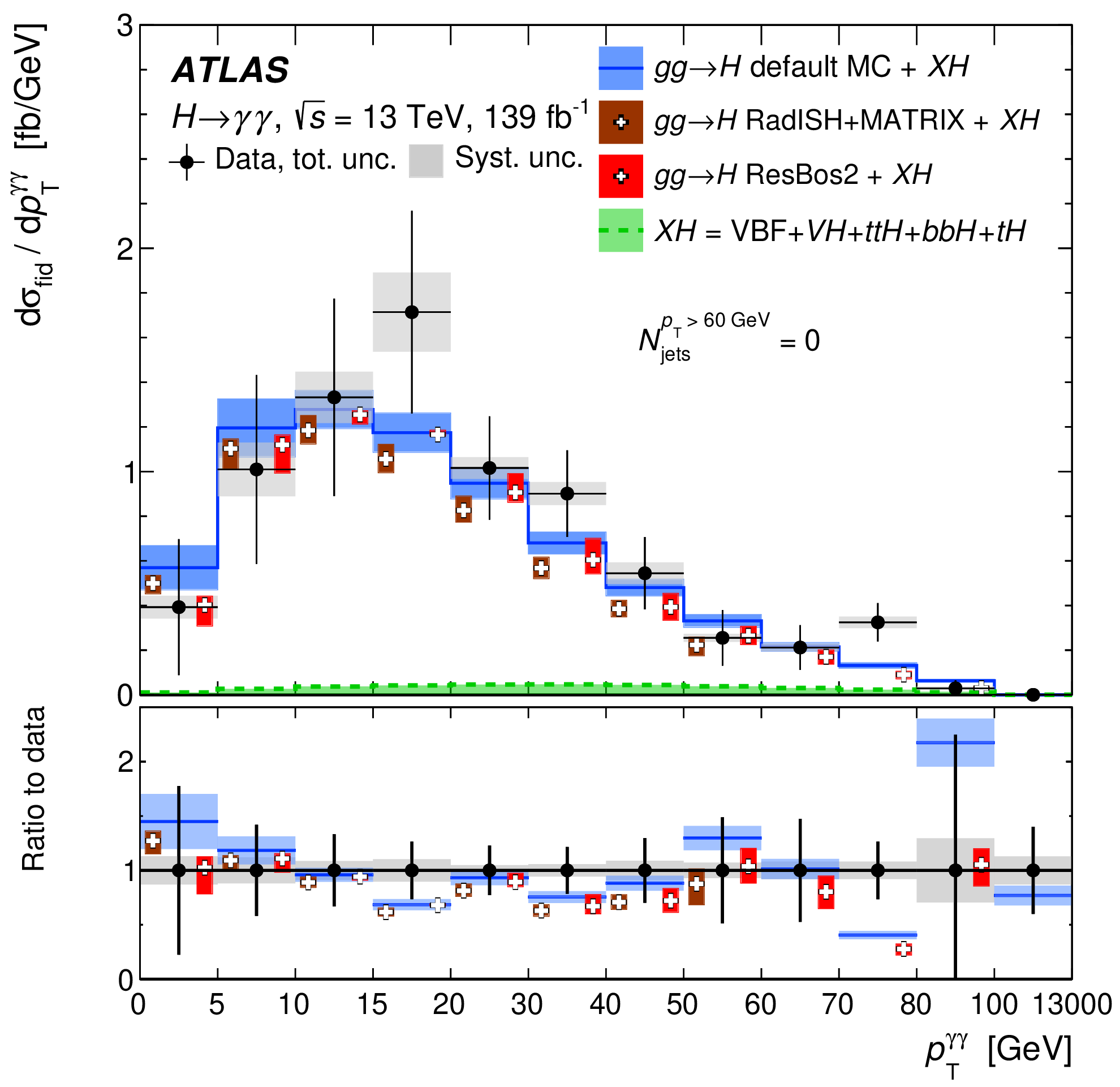}
\caption[]{A double-differential measurements of the \hgg cross section with respect to the transverse momentum of the diphoton system and the number of jets using CMS analysis~\cite{CMS:hgg}(left) and  a particle-level fiducial differential cross sections times branching ratio for \pt of \hgg with a 60 GeV jet veto using ATLAS analysis~\cite{ATLAS:hgg}(right)}
\label{fig:hgg}
\end{figure}

Figure~\ref{fig:hgg}(left) presents a double-differential measurements of the \hgg cross section with respect to the transverse momentum of the diphoton system and the number of jets by the CMS experiment.
The observed fiducial differential cross section values are shown as black points with the vertical error bars
showing the full uncertainty, the horizontal error bars show the width of the respective bin. The coloured
lines denote the predictions from different event generators. All of them have the
HX=VBF+VH+ttH component from MADGRAPH5 aMC@NLO in common, which is displayed
in violet without uncertainties. The red lines show the sum of HX and the ggH component from
MADGRAPH5 aMC@NLO reweighted to match the NNLOPS prediction.
Parametrization v.s. \pt of Higgs probes the perturbative QCD modelling of the ggH production mode. While the low \pt region is sensitive to the Yukawa coupling of the b and charm quark and QCD resummation, the high \pt region is mostly sensitive to top quark coupling, thanks to its higher mass, and BSM scenarios.
The differential cross section parametrized v.s. the jet multiplicity is also sensitive to the production mode composition and gluon emission.
The measured differential cross sectionction is in good agreement with the simulation, and is statistically limited. 
A $\chi^2$ test was used to evaluate $p$-values for the compatibility of the
measured differential cross sections and the prediction. The  $p$-values for the current measurement is found to be 0.135.

Figure~\ref{fig:hgg}(right) presents differential measurements of the \hgg cross section as a function of \pt of diphoton for events with no jet with \pt above 60 GeV with the ATLAS experiment. Similar measurements have been performed with different \pt of the vetoed jet from 60 GeV to 30 GeV. Such results provide more insight on the impact of the QCD resummation of the Higgs boson production in gluon-gluon mode.

\begin{figure}[http]
\centering
\includegraphics[width=0.35\linewidth]{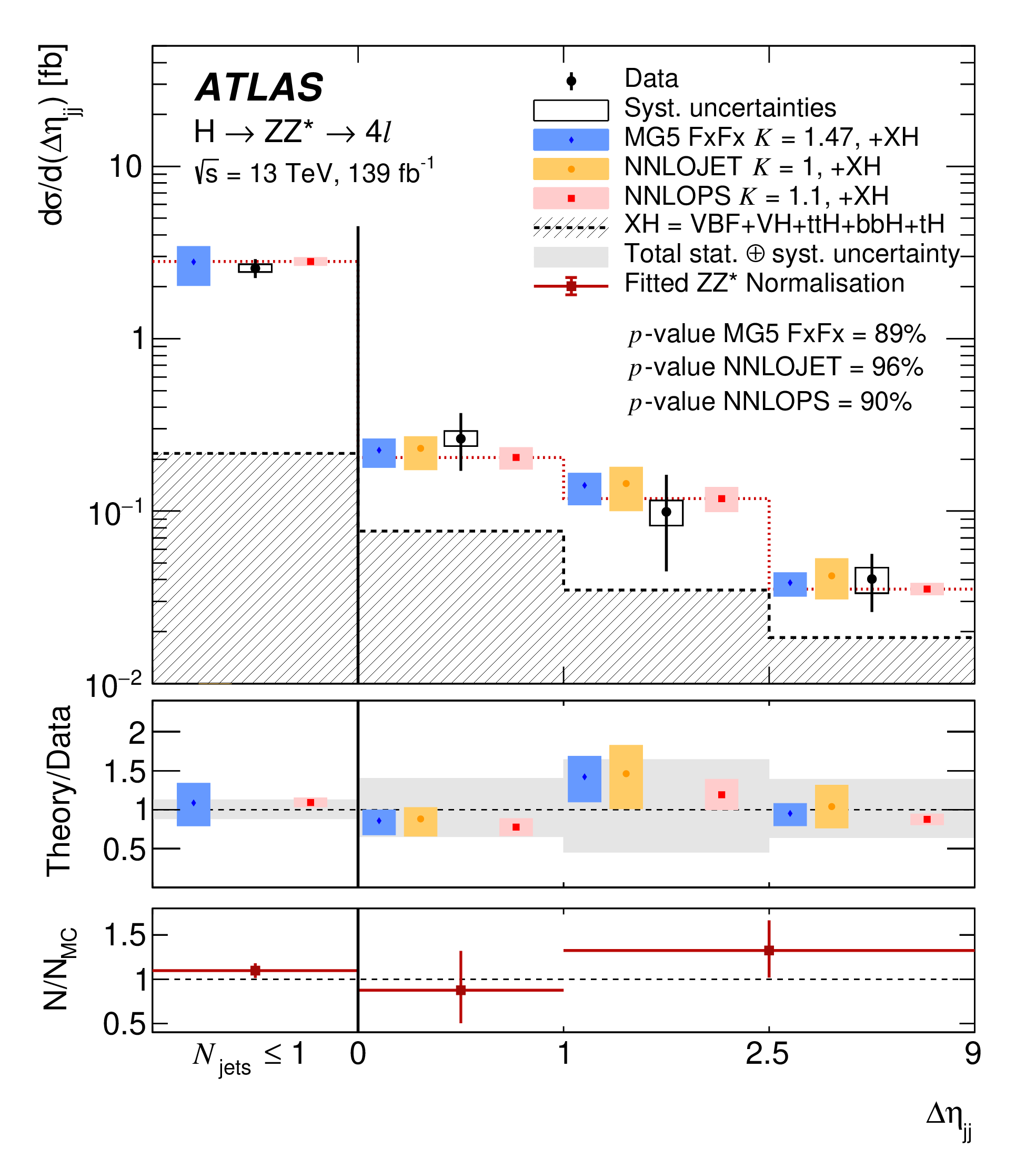}
\includegraphics[width=0.42\linewidth]{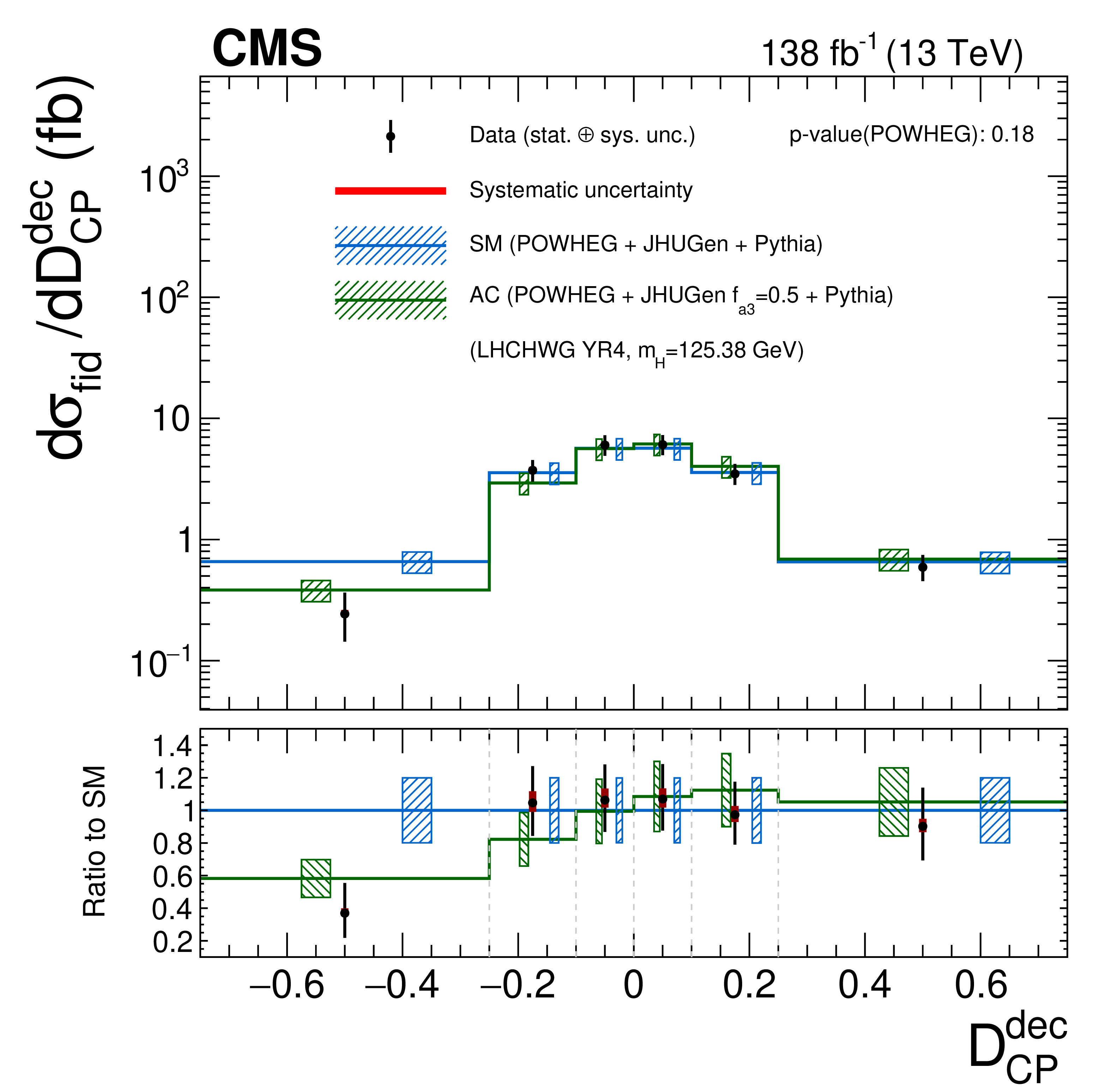}
\caption[]{Fiducial differential cross sections for the distance between the two leading-in-\pt jets in pseudorapidity, $|\Delta\eta_{jj}|$, from ATLAS analysis~\cite{ATLAS:hzz}(left) and differential cross section as functions of the matrix element kinematic discriminant $D_{CP}^{dec}$ in CMS analysis~\cite{CMS:hzz}(right)}
\label{fig:hzz}
\end{figure}

The next golden channel is \hzz where Higgs boson decay to a pair of Z bosons and each Z boson decays to a pair of muons or electrons. This is the cleanest channel among all five decay modes with the largest signal-to-background ratio in the interval around the Higgs boson mass. The major background arises from the ZZ continuum that is estimated from MC simulation and validated with data. 
The invariant mass of the four leptons is used to measure the Higgs boson differential cross section in terms of various variables. The Figure~\ref{fig:hzz} (left) shows the measured fiducial differential cross sections as a function of the pseudorapidity difference of the two leading-in-\pt jets, $\Delta\eta_{jj}$, in the ATLAS measurement. 
The experimental results are compared with theoretical predictions from various event generators. The inner boxes on the data points show the statistical uncertainties, while the total uncertainties are indicated by the outer boxes. The bottom panel shows the ratio of different predictions to the data. Such a measurement is specifically sensitive to the VBF production mode.

 The Figure~\ref{fig:hzz} (right) depicts the Higgs boson differential cross section  as a function of a matrix element kinematic discriminant $\mathrm{D_{CP}^{dec}}$ in the CMS analysis. The $\mathrm{D_{CP}^{dec}}$ is a Matrix element kinematic discriminant sensitive to the Higgs boson CP-mixing. It is calculated using the angular information of the Higgs boson, and the Z boson, and the four leptons in the events. The green histogram shows the distribution of the discriminant for the HVV anomalous coupling scenario corresponding to maximally mixing scalar and pseudoscalar scenarios. The subdominant component of the signal (VBF+VH+ttH) is fixed to the SM prediction.
Black points represent the measured fiducial cross sections in each bin. The lower panel displays the ratios of the measured cross sections and the predictions from 
POWHEG and MadGraph-5\_aMC@NLO to the NNLOPS theoretical predictions. While the SM distribution exhibits a symmetric behaviour around zero in the $\mathrm{D_{CP}^{dec}}$ spectrum, the distribution representing the anomalous coupling is asymmetric. 
The $p$-value for the compatibility of the measured differential cross section with the SM prediction is found to be 0.18.

\begin{figure}[http]
\centering
\includegraphics[width=0.52\linewidth]{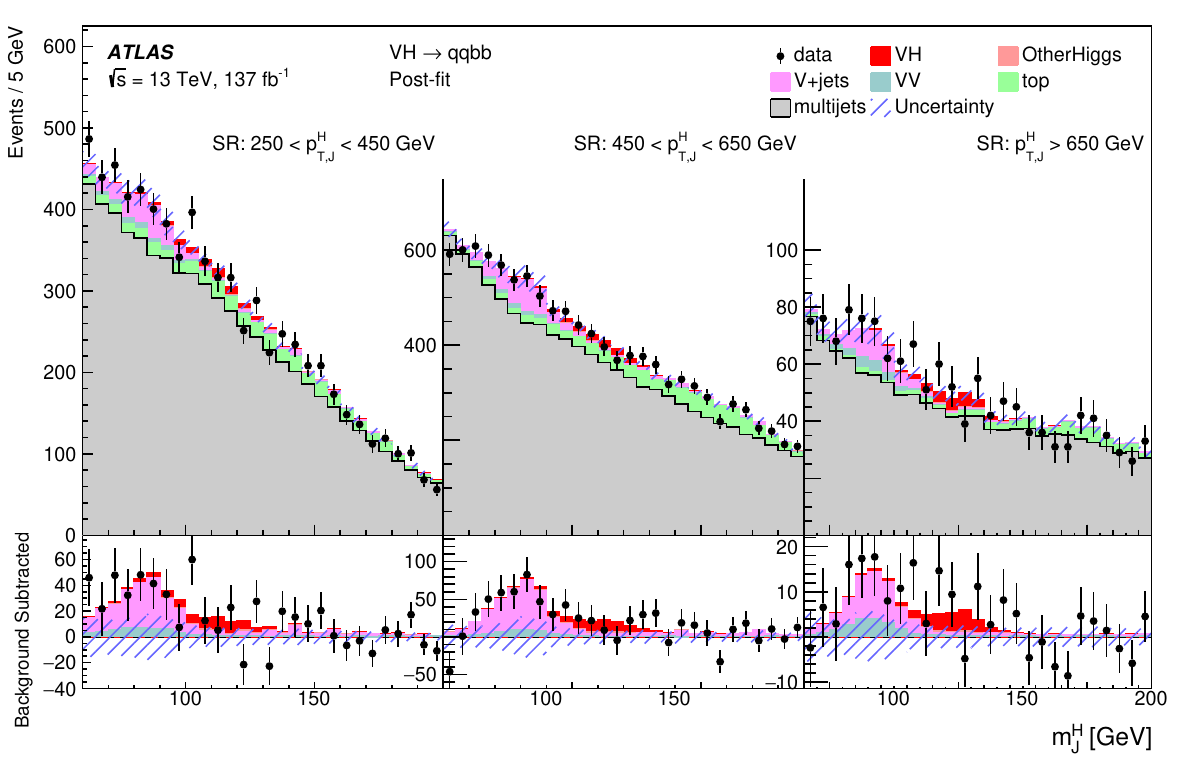}
\includegraphics[width=0.35\linewidth]{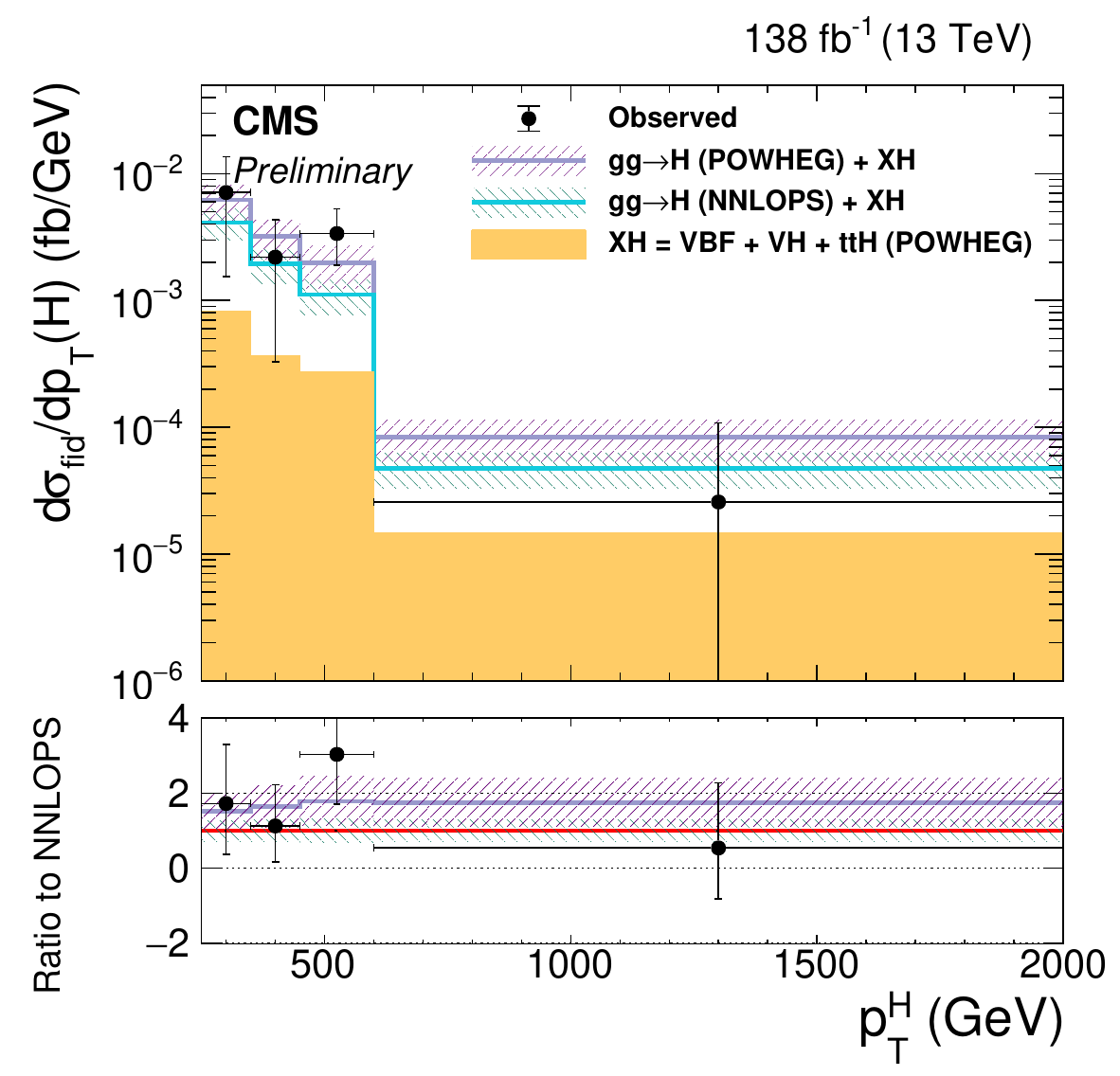}
\caption[]{The \hbb candidate jet mass distributions in the signal region for three \pt region in ATLAS analysis~\cite{ATLAS:hbbWZ}(left) and Observed and expected fiducial differential cross sections in bins of Higgs boson \pt in CMS \htt analysis~\cite{CMS:httboost}(right)}
\label{fig:hfermions}
\end{figure}

The \hbb channel represents the Higgs boson decay mode with the largest branching fraction though with a large multijet background and poor mass resolution. To distinguish the Higgs boson signal from dominant multijet background, the W/Z associated production of the Higgs boson is exploited. To select events with large signal to background ratio a highly boosted topology is employed. In such a topology the two b quarks from the Higgs boson decay will be merged into a single wide-radius jet. A dedicated tagging algorithm exploiting b-tagging properties is used to identify jets consistent with Higgs bosons decaying into bb.
The figure~\ref{fig:hfermions} (left) shows the Higgs boson candidate jet mass distributions in the signal region in various \pt ranges, as studied by the ATLAS Collaboration. The Higgs boson signal has been obtained after the inclusive fit with a single Z+jets normalization factor and a single signal strength for each \pt region. The bottom panels show the distributions after subtracting the multijet and top-quark backgrounds. The observed signal strengths in the three regions are found to be $0.8^{+2.2}_{-1.9}$, $0.4^{+1.7}_{-1.5}$, and $5.3^{+11.3}_{-3.2}$, respectively.  
 
The $\tau\tau$ channel has also been explored in the highly boosted topology. In such a topology, the standard tau lepton reconstruction efficiency loses its performance. A dedicated boosted tau reconstruction algorithm is employed to maintain the high signal selection efficiency even for regions where the Higgs boson is highly boosted and the two tau leptons into which the Higgs boson decays overlap. The dominant Drell--Yan background has been suppressed through the use of a deep neural network (NN) in this analysis. The fiducial differential production cross sections measured as a function of \pt is shown in Figure~\ref{fig:hfermions} (right). This measurement extends the probed large-\pt region beyond 600 GeV. No significant deviation with respect to the SM predictions is observed and the measurements are compatible with both the POWHEG and NNLOPS expectations. The fit has a $p$-value with respect to the SM expectation from the NNLOPS prediction of 56\%.

\section{Interpretation based on SMEFT}
\label{sec:SMEFT}

The SMEFT is a framework to explore potential new physics beyond the SM. By extending the SM Lagrangian with higher-dimensional operators, SMEFT systematically incorporates the effects of unknown heavy particles and interactions, which might be too massive to be produced directly in experiments. These higher-dimensional operators are suppressed by powers of a high energy scale, typically associated with the mass of new particles, and encode deviations from the SM predictions. 
Through global fits to data, SMEFT interpretations could constrain or reveal potential new interactions that could not otherwise be accessed directly
The effective Lagrangian can be written as the following:

\begin{equation}
    \mathcal{L}_{\mathrm{SMEFT}}=\mathcal{L}_{\mathrm{SM}}+\sum_{i=5}^{\infty} \sum_{j=0}^{N_i} \frac{c_j^{(i)}}{\Lambda^{i - 4}} O_j^{(i)}
\end{equation}

where the $c_j$ are dimensionless Wilson coefficients (WCs), the $O_j$ refer to the operators, and $\Lambda$ is the energy scale. Dimension 5 and 7 operators violate lepton and baryon number conservation and are not considered in this study. The lowest higher order operator considered in this analysis is dimension 6.

Fiducial differential cross section measurements of the Higgs boson \pt, performed by the ATLAS experiment in the \hzz and \hgg channels are used to constrain the WCs of SMEFT operators. The three most relevant operators are as follows:
\begin{itemize}
\item $O_{HG}$ : modifies the value and \pt-dependence of the ggF and \hgg partial decay width.
\item $O_{tG}$ : introduces a ttHg vertex and leads to additional contributions to the amplitude for ggF or ttH Higgs boson production,  as well as for \hgg decay.
\item $O_{tH}$ : modifies the ttH  vertex and affects Higgs boson production through top-quark-loop  mediated gg fusion and top-quark-loop amplitude.
\end{itemize}

One way to constrain the WCs is to treat them independently and set limit on one coefficient at a time and set the remaining coefficients to zero.  However, the approach adopted in the current analysis is to define a new set of eigenvectors, i.e. $ev^{[i]}$ (i = 1..3) of the Fisher information matrix that is a linear combination of the above WCs and  constrain the eigenvalues. These three eigenvectors are defined as follows:

\[
\left(
\begin{array}{ccc}
ev^{[1]} = 0.999~c_{HG} - 0.035~c_{tG}- 0.003~c_{tH} ,   \\
ev^{[2]} = 0.035~c_{HG} + 0.978~c_{tG} + 0.205~c_{tH},   \\
ev^{[3]} = -0.005~c_{HG} - 0.205~c_{tG} + 0.979~c_{tH}  \\
\end{array}
\right)
\]

\begin{figure}
\centering
\includegraphics[width=0.5\linewidth]{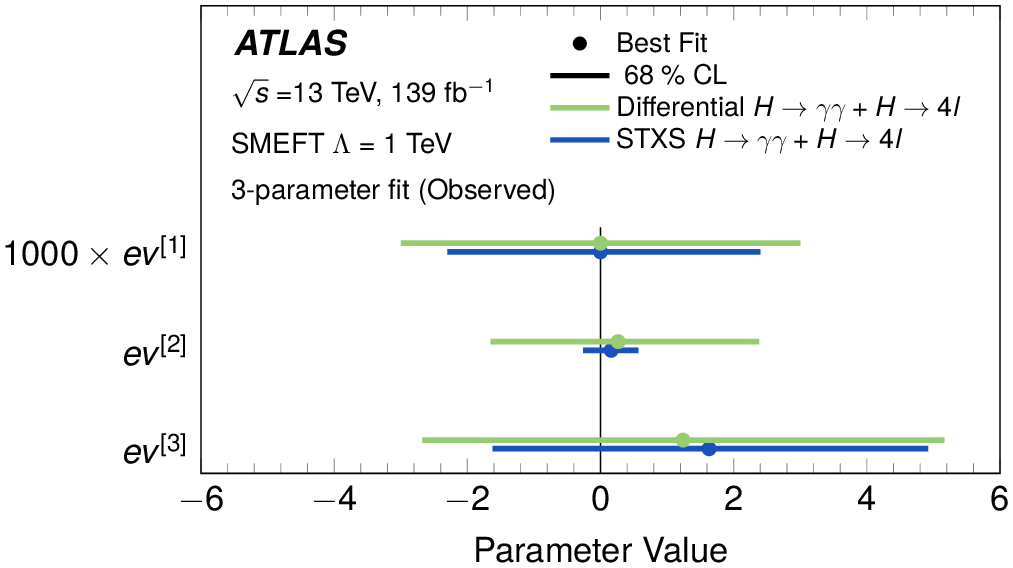}
\caption[]{The observed 68\% CL intervals on the three rotated parameters $ev^{[i]}$  obtained with the SMEFT linearised model using either Simplified Template Cross Section (blue) or fiducial \ptn-differential cross section measurements (green) in the \hgg and \hzz decay channels.}
\label{fig:SMEFT}
\end{figure}

 
 Figure~\ref{fig:SMEFT} shows the observed 68\% CL  intervals on the three rotated parameters $ev^{[i]}$ obtained with the SMEFT linearised model using either fiducial \pt-differential cross section measurements  or Simplified Template Cross Section (STXS). 
 The constraint on the first rotated parameter, which is almost
aligned with $c_{HG}$ and mainly affects the ggF production, is of the order of 1\%.  The constraints on the other two rotated parameters that are close to $c_{tG}$ and $c_{tH}$ and impact both ggF and  $t\bar{t}H$production mode, respectively is O(1). 

The current results, along with those presented in the references, show no significant deviation from SM predictions.


\end{document}